\documentclass{nsr}

\usepackage{amsmath,graphicx,array}
\usepackage{dcolumn,soul}%

\usepackage{amsthm}
\usepackage{amsmath}
\usepackage[figuresright]{rotating}%
\usepackage{algorithm, algorithmicx, algpseudocode}
\usepackage{listings}%
\usepackage{hyperref}
\usepackage{multirow}
\usepackage{booktabs}
\usepackage{tabularx}
\usepackage{subfigure}

\makeatletter
\def\uns{\ifmmode\,\else$\,$\fi}%

\makeatother
 
\jvol{00}
\jnum{nwag662}
\jyear{2026}
\doi{10.1093/nsr/nwag662}
\received{3 February Year}
\revised{23 August Year}
\accepted{3 September Year}

\begin{document}

\dhead{REVIEW}

\subhead{INFORMATION SCIENCE}

\title{Self-evolving Embodied AI}

\author{Tongtong Feng}
\author{Xin Wang$^{*}$}
\author{Wenwu Zhu$^{*}$}

\affil{Department of Computer Science and Technology, Beijing National Research Center for Information Science and Technology, Tsinghua University, Beijing 100084, China}

\authornote{\textbf{Corresponding authors.} Email: wwzhu@tsinghua.edu.cn; xin{\_}wang@tsinghua.edu.cn}

\abstract[ABSTRACT]{
Embodied Artificial Intelligence (AI) is an intelligent system formed by agents and their environment through active perception, embodied cognition, and action interaction. Existing embodied AI remains confined to human-crafted settings, in which agents are trained on given memories and construct models for given tasks, enabling fixed physical embodiments to interact with relatively static environments. Such methods fail in in-the-wild settings characterized by variable physical embodiments and dynamic open environments. This paper introduces self-evolving embodied AI, a new paradigm in which agents operate based on their changing physical embodiment and environment through five software-level components, namely memory self-updating, task self-switching, environment self-prediction, embodiment self-adaptation, and model self-evolution, together with a hardware-level component for physical embodiment evolution, aiming to achieve continually adaptive intelligence through software-hardware co-evolution. Specifically, we present the definition, framework, components, mechanisms, roadmap, and evaluation indicators of self-evolving embodied AI, systematically review state-of-the-art works for realized components, discuss practical applications, and point out future research directions. We believe that self-evolving embodied AI enables agents to autonomously learn and interact with environments in a human-like manner and provides a new perspective toward general artificial intelligence.
}


\maketitle

\section{1. Introduction}
Embodied Artificial Intelligence (AI) originates from the Embodied Turing Test by Alan Turing in 1950 \cite{turing2007computing}, which emphasizes that intelligence should be evaluated through an agent’s ability to perceive and act in the physical world in addition to abstract symbol computation. Specifically, embodied AI refers to intelligent systems \cite{feng2025embodied} formed by agents and their environment through active perception, embodied cognition, and action interaction. Unlike disembodied AI that operates on static data or abstract symbols, embodied AI emphasizes the closed-loop coupling between perception, cognition, and action \cite{sun2024comprehensive}, enabling agents to interact with physical environments in a grounded manner. As such, embodied AI benefits in promising potential in a wide range of applications, including robotics \cite{xu2024survey}, autonomous driving \cite{ma2024survey}, and other real-world interactive systems \cite{hu2023toward} etc.

\begin{figure*}[t]
    \centering
    \includegraphics[width=\textwidth]{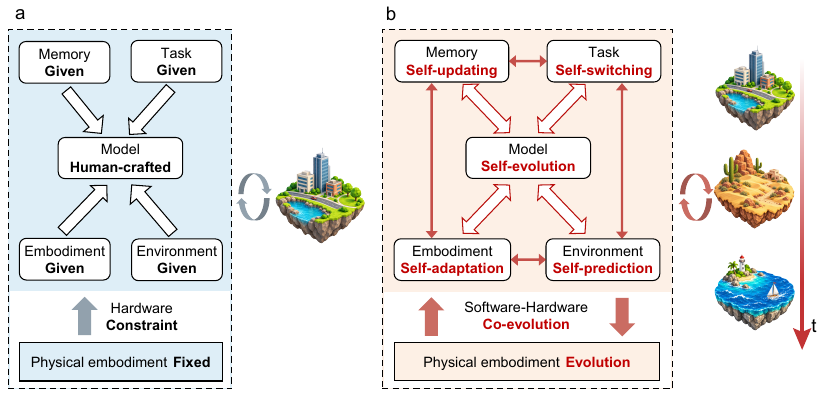}
    \caption{Comparison between existing embodied AI and self-evolving embodied AI.
    (a) Existing embodied AI operates with given memory, given task, given embodiment representation, and given environment, using a human-crafted model, while its fixed physical embodiment imposes hardware constraints. This paradigm relies heavily on external human guidance and empirical configurations.
    (b) Self-evolving embodied AI operates based on its changing physical embodiment and environment through five software-level components, namely memory self-updating, task self-switching, environment self-prediction, embodiment self-adaptation, and model self-evolution, together with a hardware-level component for physical embodiment evolution. Here, embodiment self-adaptation updates the software-level representation of the physical embodiment, whereas physical embodiment evolution changes the hardware-level embodiment itself. In this paradigm, agents can achieve continually adaptive intelligence through software-hardware co-evolution.}
    \label{fig1}
\end{figure*}

Despite its rapid progress, existing embodied AI remains confined to human-crafted setting \cite{duan2022survey, srivastava2022behavior}, in which agents are trained with given memories and construct models for given tasks, thus enabling fixed physical embodiments to interact with relatively static environments \cite{brohan2022rt1}. The existing embodied AI learning paradigm \cite{o2024open} is optimized for given tasks under predefined objectives, and the performance relies heavily on external human guidance and empirical specification \cite{zitkovich2023rt}, including human-given tasks, predesigned models, manually-collected datasets, fixed-configuration physical embodiments, and specified experimental environments.

However, real-world scenarios often involve in-the-wild setting. On the one hand, in-the-wild environments are often dynamic and open \cite{hafner2025mastering, bruce2024genie}, with continually changing objects, dynamics, and interaction patterns that cannot be fully predefined or exhaustively enumerated during training. On the other hand, in-the-wild deployment often requires variable physical embodiments \cite{liu2025aligning}, where agents differ in morphology, sensing, actuation, computational configurations, and physical constraints, resulting in models learned under fixed physical embodiments being difficult to adapt. Unfortunately, existing embodied AI fails in in-the-wild setting.

Motivated by neuroscience \cite{zhu2024advancing} and cognitive science \cite{wang2024comprehensive}, we introduce a new paradigm of 'self-evolving embodied AI', as illustrated in Fig. \ref{fig1}. Self-evolving embodied AI requires agents to operate based on their changing physical embodiment and environment through five software-level components, namely memory self-updating, task self-switching, environment self-prediction, embodiment self-adaptation, and model self-evolution, together with a hardware-level component for physical embodiment evolution. Here, embodiment self-adaptation updates the software-level representation of the physical embodiment, whereas physical embodiment evolution changes the hardware-level embodiment itself. Software evolution guides physical embodiment evolution, which in turn reshapes software evolution. In this paradigm, agents can achieve continually adaptive intelligence through software-hardware co-evolution, enabling embodied AI to move from human-crafted setting to in-the-wild setting.

Specifically, we first present the definition, framework, components, mechanisms, roadmap, and evaluation indicators of self-evolving embodied AI. We then systematically review state-of-the-art works for each self-evolving component. Furthermore, we discuss practical applications that demonstrate the advantages of self-evolving embodied AI in variable physical embodiments and dynamic open environments. Finally, we summarize future research directions toward building controllable, trustworthy, and swarm self-evolving embodied AI. 

It should be noted that self-evolving embodied AI is currently a vision-guided paradigm rather than a review of a mature technical system. No existing system has yet realized full-stack autonomous co-evolution between all five software-level components and a hardware-level component. Existing studies should be regarded as partial realizations that provide preliminary building blocks toward the ideal full-loop mechanism. Nevertheless, we anticipate with full confidence that self-evolving embodied AI is able to learn autonomously in a human-like manner and provide a new perspective toward general artificial intelligence.

\begin{table*}[t]
\caption{Comparison between adaptive intelligence and self-evolving intelligence.}
\label{tab:ad}
\begin{tabular}{>{\raggedright\arraybackslash}p{0.14\textwidth} >{\raggedright\arraybackslash}p{0.18\textwidth} >{\raggedright\arraybackslash}p{0.40\textwidth} >{\raggedright\arraybackslash}p{0.18\textwidth} }
\toprule
Type
& Level of Change
& Object of Change 
& Goal of Change \\
\midrule
Adaptive
& Behavior-level adjustment within a given system
& Model parameters
& Adaptive capabilities\\
\midrule
Self-evolving 
& System-level self-correction and self-improvement
& Memory organization, task formulation, world modeling, embodiment representation, model architecture, and physical embodiment itself
& New or improved capabilities, structures, and mechanisms\\
\bottomrule
\end{tabular}
\end{table*}

\section{2. Self-evolving Embodied AI}
\label{sec:sei}
This section introduces 'self-evolving embodied AI', a new paradigm that aims to move embodied AI beyond human-crafted setting toward in-the-wild setting. We first provide a precise definition and unified framework and then elaborate on its five software-level components and hardware-level component. We next explain two coupled self-evolving mechanisms, namely co-evolution among software-level components and software-hardware co-evolution. Building on these mechanisms, we formulate a closed-loop roadmap for component triggering, software-hardware co-evolution, constraint verification, and continual state transition. Finally, we introduce evaluation indicators for measuring component-level and system-level self-evolution.

\subsection{2.1 The Definition and Framework}
\label{subsec:definition}
This subsection establishes the conceptual foundation of self-evolving embodied AI. We provide a precise definition of self-evolving embodied AI and then introduce a unified framework to describe how self-evolution is organized.

\textbf{Definition 1} (Self-evolving embodied AI). An agent operates based on its changing physical embodiment and external environment through five software-level components, namely memory self-updating, task self-switching, environment self-prediction, embodiment self-adaptation, and model self-evolution, together with a hardware-level component for physical embodiment evolution, all of which are coupled through software-hardware co-evolution to achieve continually adaptive intelligence.

As shown in Table \ref{tab:ad}, adaptive intelligence \cite{charlesworth2024human} and self-evolving intelligence represent two different intelligence types in embodied scenarios. Adaptive intelligence focuses on post-deployment behavioral adjustment under changing conditions, such as shifts in data distributions, task requirements, or environmental states. It changes the agent’s behavior within a given agent system. Self-evolving intelligence, however, refers to feedback-driven self-correction and self-improvement of the agent system itself. The key distinction is that adaptive intelligence modifies the behavior produced by a given system, whereas self-evolving intelligence modifies the system that produces behavior. In self-evolving embodied AI, the object of change includes memory organization, task formulation, world modeling, embodiment representation, model architecture, and the physical embodiment itself.

\paragraph{Framework} Figure~\ref{fig1}b illustrates the framework of self-evolving embodied AI. Specifically, the software layer is organized around five tightly coupled modules, namely memory self-updating, task self-switching, environment self-prediction, embodiment self-adaptation, and model self-evolution. At the hardware level, physical embodiment evolution enables changes to the agent’s physical embodiment itself. The five software-level modules are not independent, but evolve jointly through continuous interaction with the environment and with each other. Updates in one module may induce adjustments in others. This organization is reflected by bidirectional information exchange among modules across time. For example, changes in physical embodiment constraints may alter the relevance of stored memory, which in turn affects task switching and model evolution. Software evolution can also guide physical embodiment evolution, which in turn reshapes software evolution. Together, the software-level components and the hardware-level component form a unified evolutionary loop that supports continually adaptive intelligence under variable physical embodiments and dynamic open environments.

For comparison, Fig.~\ref{fig1}a summarizes the framework of existing embodied AI \cite{Wang2026-1, feng2025embodied, liu2025aligning, xu2024survey, ma2024survey, duan2022survey}. Specifically, the agent with given physical embodiment, memory, task, uses a human-crafted model for predefined objectives in a specific environment. This framework is largely unidirectional, shown as one-way arrows pointing toward the human-crafted model.

The contrast between the two frameworks highlights a fundamental difference in how embodied AI is realized. Existing embodied AI relies on given components and human-crafted models with external human guidance and empirical configurations, whereas self-evolving embodied AI organizes memory self-updating, task self-switching, environment self-prediction, embodiment self-adaptation, and model self-evolution as five software-level components, while these components co-evolve with a hardware-level component for physical embodiment evolution. Compared to existing embodied AI operating under relatively static environments, by replacing unidirectional optimization with bidirectional coupling among self-evolving modules and between the software and hardware levels, self-evolving embodied AI enables continual adaptation with autonomous evolution in dynamic open environments.

Recent studies have shown a growing interest in self-evolving agents \cite{fang2025comprehensive, gao2025survey, Wang2026-6, Wang2026-4, Wang2026-2, Wang2026-3}, which aim to improve adaptability through continual interaction with the environment. Specifically, those methods explore self-evolution by using prompt engineering \cite{ou2025symbolic, yadav2025adaptive}, memory selection \cite{cai2025building, xu2025sedm, zhang2025memgen, huang2025r}, and tool refinement technologies \cite{liu2025agent0}, and are often built upon pretrained Large Language Models (LLMs) as unified end-to-end backbones \cite{wang2025mobile, thawakar2025evolmm}. Those methods demonstrate the potential of self-evolution without explicit human intervention, but environment perception blocks and physical embodiment configurations are typically treated as fixed once deployed \cite{zhang2025evolvesearch, sun2025seagent}, limiting their ability to support long-term autonomy under variable physical embodiments and dynamic environments. 

In summary, to the best of our knowledge, this paper is the first systematic formalization of self-evolution as a top-level paradigm for embodied AI. Specifically, self-evolving embodied AI arises from closed-loop evolution among five software-level components and their bidirectional co-evolution with a hardware-level component, driven by the agent’s self-state transitions by action-feedback interactions.

\subsection{2.2 The Components}
\label{subsec:components}
This subsection elaborates on the core components of self-evolving embodied AI. For each components, we explain what evolves, why evolution is necessary, when evolution is triggered, how evolution is realized.

\paragraph{Memory self-updating} What evolves in memory self-updating is the agent’s internal memory representation. Experience is continuously generated through the agent's interaction with the environment, while memory self-updating determines how experience is selectively retained, revised, or discarded over time. Why evolution is necessary lies in the fact that fixed memories or static datasets cannot support long-term adaptation under environmental distribution shift and changing physical embodiments or tasks. In in-the-wild environments, storing all past experiences is neither feasible nor desirable. When evolution is triggered, previously stored experiences become outdated, irrelevant, or misleading with respect to the agent’s current physical embodiment constraints, task objectives, or environmental dynamics. How evolution is realized is through selective memory updating mechanisms, such as memory self-editing, memory self-organization, and memory self-distillation, which prioritize experiences based on relevance, novelty, uncertainty, or long-term utility. This component has relatively high feasibility, as recent agent-memory studies \cite{xu2025amem, liang2024self, zhao2024expel} have already explored memory editing, organization, and distillation for long-horizon interaction. However, key bottlenecks remain in long-term memory reliability, conflict resolution among inconsistent experiences, privacy of stored interaction histories, and preventing outdated memories from misleading future decisions.

\paragraph{Task self-switching} What evolves in task self-switching is the agent’s internal representation of task objectives. Rather than optimizing predefined objectives of given tasks, task self-switching improves the agent’s ability to autonomously adjust what it is trying to achieve over time. Why evolution is necessary lies in the fact that predefined objectives of given tasks cannot adequately capture the changing goals, constraints, and opportunities encountered in dynamic open environments. In-the-wild setting rarely present stable, well-defined goals; over time, predefined objectives of given tasks may limit autonomy and prevent effective adaptation. When evolution is triggered, current task formulations become infeasible, suboptimal, or misaligned with the agent’s internal state, physical embodiment constraints, or environmental dynamics. How evolution is realized is through autonomous task self-switching mechanisms, such as task self-selection and task self-generation, enabling agents to continuously adjust what they aim to achieve without explicit human re-specification. This component is scientifically reasonable \cite{chen2025self, yang2025zerogui}, but its engineering feasibility is more challenging than memory updating. The main bottlenecks lie in goal drift, task misalignment, unsafe task generation, and the lack of stable constraints for deciding when an agent should switch, decompose, or generate tasks autonomously.

\paragraph{Environment self-prediction} What evolves in environment self-prediction is the agent’s internal representation of the external world. It captures how agents continuously update their understanding of environmental dynamics and make future predictions of the external world. Why evolution is necessary lies in the fact that real-world environments are non-stationary and cannot be fully characterized by fixed or offline-learned models. Agents need to maintain and refine predictive representations that anticipate future states, rewards, and action-conditioned interaction sequences. When evolution is triggered, newly observed environment states contradict prior predictions, reveal previously unseen dynamics, or expose long-term dependencies that were not captured before. How evolution is realized is through continual refinement of world models based on ongoing interaction, such as understanding world models, generative world models, and collaborative world models, allowing agents to update their internal representation of the external world and supporting adaptive planning and decision-making over time. This component also has relatively high feasibility because world models \cite{hafner2025mastering, bruce2024genie, Wang2026-5, liu2025tcdformer} have shown strong potential in representation learning, imagination, and model-based planning. Nevertheless, practical bottlenecks remain in long-horizon rollout consistency, uncertainty calibration, causal prediction, and action-conditioned generalization under dynamic open environments.

\paragraph{Embodiment self-adaptation} What evolves in embodiment self-adaptation is the agent’s internal representation of its own physical state. Self-evolving embodied AI needs to adapt to changes in the physical embodiment’s heterogeneous morphology, sensing capabilities, actuation limits, computational configurations, and physical constraints. Why evolution is necessary lies in the fact that embodiments may vary across platforms or change over time due to reconfiguration, causing policies learned under fixed physical embodiments to fail. When evolution is triggered, discrepancies arise between expected and actual physical embodiment states, indicating that previous embodiment assumptions are no longer valid. How evolution is realized is through software-level embodiment-aware adaptation, such as embodiment self-transfer, embodiment self-calibration and embodiment self-recovery, enabling agents to maintain functionality across variable physical embodiments. This component is feasible as a partial function \cite{patel2025get, shen2026metaworld, yu2017preparing, wang2025improving}, especially in cross-embodiment transfer, online calibration, and damage recovery. However, reliable deployment remains difficult due to sim-to-real gaps, and the need to ensure physical feasibility under morphology, sensing, actuation, or energy changes.

\paragraph{Model self-evolution} What evolves in model self-evolution is the agent’s internal model design, including model architectures, optimization strategies, and evaluation criteria, rather than only model parameters. Why evolution is necessary lies in the fact that fixed architectures, given training strategies, and predefined evaluation metrics cannot remain effective under long-term changes in memory, tasks, environments, and physical embodiments. As operating conditions evolve, models that were previously well-optimized may become inefficient, misaligned, or inadequate. When evolution is triggered, persistent performance degradation, increasing uncertainty, or systematic mismatch between evaluation outcomes and real-world behavior indicates that current model designs are no longer appropriate. How evolution is realized is through adaptive modification of model architectures, adjustment of optimization strategies, and refinement of evaluation criteria, such as model self-restructuring, model self-optimization, and model self-evaluation, enabling the learning process itself to evolve and remain aligned with long-horizon autonomous operation. Compared with the other components, model self-evolution is still the most under-explored and least mature. Existing works \cite{yan2024masked, kang2025self, zheng2023judging} provide preliminary evidence for model restructuring, self-optimization, and self-evaluation, but most of them remain limited to modular routing, adapter composition, self-refinement, or LLM-based judging. Critical bottlenecks include insufficient theoretical guarantees for convergence, stability, safety, and alignment, as well as limited empirical support for autonomous model evolution in embodied closed-loop systems.

\paragraph{Physical embodiment evolution} Physical embodiment evolution encompasses reconfiguration within a predefined hardware architecture and design space, as well as transformation that alters or expands the hardware architecture or design space itself. Reconfiguration changes modules, sensors, or actuators while preserving the underlying architecture, whereas transformation may fundamentally change morphology, actuation mechanisms, or computing architecture, such as a transition from a fish-inspired to a mammal-inspired embodiment or from a conventional von Neumann architecture to a compute-in-memory (CIM) architecture. It therefore permits previously unseen hardware architectures and embodiment paradigms to emerge. Physical embodiment evolution is necessary when a fixed physical embodiment constrains the agent’s functional capabilities or adaptation to new tasks and environments. It is triggered when software-level adaptation alone cannot resolve a mismatch between the capabilities of the current physical embodiment and the requirements of present or anticipated tasks and environments. It is realized through software-hardware co-evolution, in which software guides changes to the physical embodiment, while the resulting changes reshape perception and action capabilities and thereby provide new feedback for software evolution. Current realizations mainly take the form of morphological reconfiguration~\cite{zhao2025modular, belke2023morphological} and functional reconfiguration~\cite{sun2023embedded}, whereas transformation remains largely unrealized in current embodied AI systems because fundamentally new hardware architectures may require lengthy design, fabrication, and validation cycles.

\subsection{2.3 Self-evolving Mechanisms}
\label{subsec:mechanisms}
Self-evolving embodied AI operates through two coupled mechanisms: co-evolution among software-level components and software-hardware co-evolution.

\paragraph{Co-evolution among software-level components} Memory self-updating, task self-switching, environment self-prediction, embodiment self-adaptation, and model self-evolution do not evolve independently. Instead, changes in one component propagate to the others through action-feedback interactions. For example, embodiment self-adaptation directly affects the entire self-evolving loop. When the agent detects a mismatch between expected and actual physical embodiment states, such as sensor degradation, actuator failure, or morphology change, this mismatch first triggers environment self-prediction to revise body-environment dynamics. It then guides memory self-updating to prioritize experiences relevant to the current physical embodiment, constrains task self-switching by filtering infeasible objectives, and further drives model self-evolution to adjust architectures or optimization strategies.

\begin{figure*}[t]
    \centering
    \includegraphics[width=\textwidth]{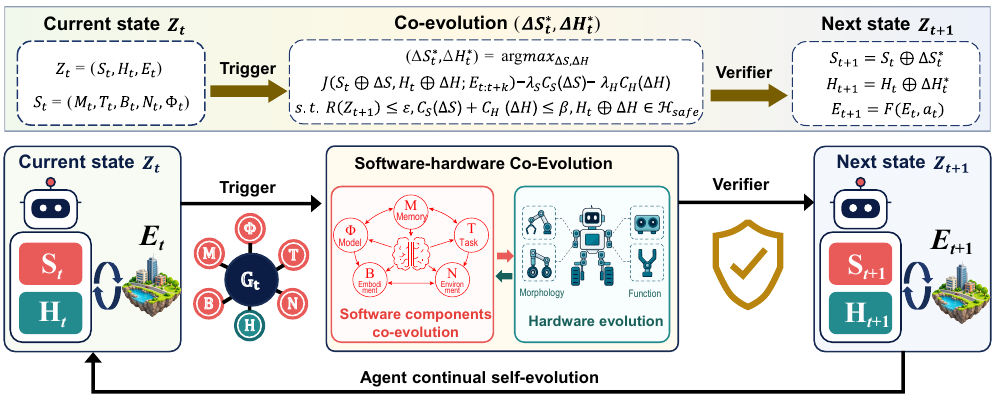}
    \caption{A roadmap for self-evolving embodied AI. At time $t$, the joint state $Z_t$ comprises the software state $S_t$, hardware state $H_t$, and external environment state $E_t$. Trigger gates $G_t$ activate the required software components and hardware component. The selected software and hardware updates are jointly optimized under performance, cost, risk, and safety constraints and are verified before being deployed to form the next state $Z_{t+1}$. Repeating this process enables agent continual self-evolution.}
    \label{fig:roadmap}
\end{figure*}

\paragraph{Software-hardware co-evolution} When software-level adaptation alone cannot satisfy the agent’s operating requirements, a mismatch between current hardware capabilities and task or environmental demands may trigger physical embodiment evolution. Depending on whether the required hardware changes remain within or extend beyond the existing hardware architecture and design space, physical embodiment evolution may be realized through reconfiguration or transformation, respectively. Software and hardware subsequently co-evolve, with changes in the physical embodiment reshaping perception and action capabilities and requiring corresponding software adaptation. As shown in Fig. 1b, this bidirectional interaction forms the software-hardware co-evolution loop. Repeated operation of this loop enables software updates and physical embodiment changes to remain coordinated as task and environmental requirements evolve.

Together, the two mechanisms enable continually adaptive intelligence with autonomous evolution in variable physical embodiments and dynamic open environments.

\subsection{2.4 Self-evolving Roadmap}
\label{subsec:roadmap}

Figure~\ref{fig:roadmap} presents a closed-loop roadmap consisting of state representation, component triggering, software-hardware co-evolution, constraint verification, and state transition. At time $t$, the joint state is defined as:
\begin{equation}
    Z_t=(S_t,H_t,E_t),
\end{equation}
\begin{equation}
    S_t=(M_t,T_t,B_t,N_t,\Phi_t),
\end{equation}
where $Z_t$ is the joint state, $S_t$ is the software state, $H_t$ is the hardware configuration, and $E_t$ is the external environment. Within $S_t$, $M_t, T_t, B_t, N_t, \Phi_t$ denote memory, task, embodiment representation, environment representation, and model, respectively. In particular, $B_t$ describes software adaptation to the physical embodiment, whereas $H_t$ describes the physical hardware itself.

The trigger gates $G_t$ select the software components and the hardware component that need to be activated. The selected software change $\Delta S$ and hardware change $\Delta H$ are jointly optimized:
\begin{equation}
\begin{aligned}
(\Delta S_t^{*},\Delta H_t^{*})
=&
\operatorname*{arg\,max}_{\Delta S,\Delta H}\\
&[J\!\left(
    S_t\oplus\Delta S,
    H_t\oplus\Delta H;
    E_{t:t+k}
\right)\\
&-\lambda_S C_S(\Delta S) -\lambda_H C_H(\Delta H)],
\end{aligned}
\end{equation}
\begin{equation}
\begin{aligned}
\text{s.t.} \quad \quad \quad \quad \quad R(Z_{t+1})
&\leq \epsilon,\\
C_S(\Delta S)+C_H(\Delta H)
&\leq \beta,\\
H_t\oplus\Delta H
&\in\mathcal{H}_{\mathrm{safe}},
\end{aligned}
\end{equation}
where $\Delta S_t^{*}$ and $\Delta H_t^{*}$ are the optimal software and hardware updates; $E_{t:t+k}$ denotes the environment over a prediction horizon of length $k$; $J$ measures the expected performance; $C_S$ and $C_H$ measure software and hardware update costs; and $\lambda_S$ and $\lambda_H$ are their corresponding weights. The operator $\oplus$ represents an implementation-dependent update. Moreover, $R$ measures risk, $\epsilon$ is the maximum acceptable risk, $\beta$ is the total update budget, and $\mathcal{H}_{\mathrm{safe}}$ is the set of safe hardware configurations.

The five software components evolve collaboratively, while software evolution guides physical embodiment evolution and the evolved physical embodiment changes subsequent perception and action. A verifier checks the constraints before the updates are deployed. Once verified, the state transition is:
\begin{equation}
\begin{aligned}
    S_{t+1} &= S_t\oplus\Delta S_t^{*},\\
    H_{t+1} &= H_t\oplus\Delta H_t^{*},\\
    E_{t+1} &= F(E_t,a_t),
    \label{eq:roadmap_transition}
\end{aligned}
\end{equation}
where $S_{t+1}$, $H_{t+1}$, and $E_{t+1}$ are the next software, hardware, and environment states; $a_t$ is the agent action; and $F$ is the environment transition function. The resulting $Z_{t+1}$ enters the next cycle, enabling agent continual self-evolution.

\subsection{2.5 Evaluation Indicators}
Evaluation indicators for self-evolving embodied AI should measure whether an agent can continuously evolve its software-level and hardware-level components, and evolve the overall agent system through long-horizon interaction. For memory self-updating, evaluation can focus on the effectiveness of memory self-editing, self-organization, and self-distillation, including whether the agent can update, structure, and abstract experiences for later decision-making~\cite{xu2025amem,yan2025memory}. For task self-switching, indicators should assess task self-selection and task self-generation, measuring whether the agent can select feasible tasks or generate new tasks that promote continuous improvement~\cite{chen2025self,fang2025webevolver}. For environment self-prediction, evaluation should consider understanding world models, generative world models, and collaborative world models, including the accuracy, consistency, and decision usefulness of latent prediction, future imagination, and world-model collaboration~\cite{hafner2025mastering,bruce2024genie,yoo2025world}. For embodiment self-adaptation, indicators should correspond to self-transfer, self-calibration and self-recovery, measuring calibration accuracy and recovery performance under morphology, sensing, actuation, or physical-constraint changes~\cite{patel2025get,ying2024peac,allard2023online}. For model self-evolution, evaluation should cover model self-restructuring, self-optimization, and self-evaluation, measuring whether architecture, optimization, and evaluation mechanisms can be autonomously improved without unstable degradation~\cite{yu2024boosting,madaan2023self,zheng2023judging}. For physical embodiment evolution, evaluation should cover morphological reconfiguration and functional reconfiguration, measuring whether morphology and function can be safely reconfigured in response to software guidance, as well as whether the underlying architecture of the agent’s physical embodiment is altered or its design space is expanded.

At the system level, self-evolving embodied AI can be evaluated by software-hardware co-evolution consistency, autonomous evolution cost, physical embodiment evolution cost, and long-horizon robustness~\cite{fang2025comprehensive,gao2025survey,cai2025building}. Co-evolution gain can further quantify whether coupling software evolution with physical embodiment evolution outperforms software-only evolution or physical embodiment evolution alone.

\begin{table*}[t]
\centering
\caption{Taxonomy of representative works for self-evolving embodied AI.}
\label{tab:taxonomy}
\renewcommand{\arraystretch}{1.1}
\setlength{\tabcolsep}{6pt}
\begin{tabular}{p{2.1cm} p{2.5cm} p{8.2cm}}
\toprule
Module & Taxonomy & Representative Works \\
\midrule
\multirow{4}{*}{\parbox{2.3cm}{Memory \\self-updating}}
& \multirow{1}{*}{Self-editing} 
& SAGE~\cite{liang2024self}, Mem0~\cite{chhikara2025mem0}, DRC~\cite{guo2024towards}, 
Memento~\cite{zhou2025memento}\\

& \multirow{2}{*}{Self-organization}
& A-MEM~\cite{xu2025amem}, MemInsight~\cite{salama2025meminsight}, 
MemGen~\cite{zhang2025memgen}, ReMe~\cite{cao2025remember}, Generative Agents~\cite{park2023generative} \\

& \multirow{1}{*}{Self-distillation}
& ExpeL~\cite{zhao2024expel}, AWM~\cite{wang2024awm}, MUSE~\cite{yang2025learning} \\

\midrule
\multirow{2}{*}{\parbox{2.3cm}{Task \\ Self-switching}}
& \multirow{1}{*}{Self-selection}
& SEC~\cite{chen2025self}, WebRL~\cite{qi2024webrl}, Agent0~\cite{xia2025agent0}, Mobile-Agent-E~\cite{wang2025mobile} \\

& \multirow{1}{*}{Self-generation}
& ZeroGUI~\cite{yang2025zerogui}, AgentEvolver~\cite{zhai2025agentevolver}, WebEvolver~\cite{fang2025webevolver} \\

\midrule
\multirow{4}{*}{\parbox{2.3cm}{Environment \\Self-prediction}}
& \multirow{2}{*}{Understanding WM}
& DreamerV3~\cite{hafner2025mastering}, DreamerV4~\cite{hafner2025training}, JEPA~\cite{assran2023self}, V-JEPA~\cite{assran2025v}, EvolvingAgent~\cite{feng2025evoagent}, ModularAgent~\cite{zhan2026modularagent}\\

& \multirow{2}{*}{Generative WM}
& Genie~\cite{bruce2024genie}, Genie-2~\cite{parker2024genie},
Matrix-Game-2~\cite{he2025matrix},
MineWorld~\cite{guo2025mineworld},
MineDreamer~\cite{zhou2024minedreamer}, OA~\cite{liu2025continual} \\

& \multirow{1}{*}{Collaborative WM}
& WorMI~\cite{yoo2025world}, COMBO~\cite{zhang2025combo}, WebEvolver~\cite{fang2025webevolver}, WKM~\cite{qiao2024agent} \\

\midrule
\multirow{3}{*}{\parbox{2.3cm}{Embodiment \\Self-adaptation}}
& \multirow{1}{*}{Self-transfer}
& GET-Zero~\cite{patel2025get}, BoT~\cite{sferrazza2024body}, PEAC~\cite{ying2024peac} \\

& \multirow{1}{*}{Self-calibration}
& UP-OSI~\cite{yu2017preparing}, SPI-Active~\cite{sobanbabu2025sampling}, OFCI~\cite{kim2025online} \\

& \multirow{1}{*}{Self-recovery}
& Fall Recovery~\cite{yang2023learning}, Damage Recovery~\cite{allard2023online} \\

\midrule
\multirow{6}{*}{\parbox{2cm}{Model \\Self-evolution}}
& \multirow{2}{*}{Self-restructuring}
& MaskTAS~\cite{yan2024masked},
PMoE~\cite{jung2024pmoe},
MoE-Adapters4CL~\cite{yu2024boosting},
D-MoLE~\cite{ge2025dynamic},
L2R~\cite{araujo2024learning},
EEP~\cite{liu2024efficient} \\

& \multirow{2}{*}{Self-optimization}
& MoE-CL~\cite{kang2025self},
SRT~\cite{hu2024teaching},
SEAS~\cite{diao2025seas},
TTRL~\cite{zuo2025ttrl},
Self-Refine~\cite{madaan2023self},
Reflexion~\cite{shinn2023reflexion} \\

& \multirow{2}{*}{Self-evaluation}
& LLM-as-a-Judge~\cite{zheng2023judging},
SER~\cite{huang2024self},
RLME~\cite{rentschler2026rlme},
RLIF~\cite{zhao2025learning},
MAE~\cite{chen2025multi}\\

\midrule
\multirow{3}{*}{\parbox{2cm}{Physical\\ Embodiment\\ Evolution}}
& \multirow{2}{*}{Morphology}
& EmbeddedSM~\cite{sun2023embedded},
PhysicalPM~\cite{belke2023morphological}, HEXEL~\cite{yoder2024hexagonal}, SCTB~\cite{zhao2025modular} \\

& \multirow{1}{*}{Function}
& M4~\cite{sihite2023multi},
UniCore~\cite{li2025reciprocal},
FlexiFly~\cite{zhao2025flexifly}
\\

\bottomrule
\end{tabular}
\end{table*}

\section{3. Methodologies}
\label{sec:methods}
Self-evolving embodied AI comprises five software-level components and a hardware-level component. As shown in Table \ref{tab:taxonomy}, existing methods realize only partial components, and no system has yet achieved closed-loop co-evolution across both software and hardware levels.

\subsection{3.1 Memory Self-updating}
\label{sec:exp_self_selection}
Existing works on memory self-updating can be broadly categorized into three classes: memory self-editing, memory self-organization, and memory self-distillation. These categories primarily differ in how experience is modified, structured, and abstracted to support long-horizon adaptation over time.

\paragraph{Memory self-editing}
Those works focus on explicit operations that directly modify stored memory content, including addition, update, deletion, merging, and forgetting. SAGE~\cite{liang2024self} introduces reflective mechanisms that regulate memory retention using forgetting curves. Mem0~\cite{chhikara2025mem0} formalizes memory editing by extracting salient facts from interaction histories and consolidating them into long-term memory. DRC~\cite{guo2024towards} establishes a closed-loop dual regression framework to tackle ill-posed super-resolution within a constrained search space, effectively refining degraded inputs and providing higher-quality experience for downstream tasks. Memento~\cite{zhou2025memento} optimizes memory rewriting and retrieval policies to maintain compact and relevant long-term memory. Together, these works treat memory evolution as an operation-driven editing process.

\paragraph{Memory self-organization}
Those works emphasize evolving the structure of memory rather than the content. A-MEM~\cite{xu2025amem} organizes agent memory as a dynamically indexed and linked note network inspired by Zettelkasten-style knowledge management. MemInsight~\cite{salama2025meminsight} augments raw episodic memory with semantic structure, enabling more robust retrieval and reuse. MemGen~\cite{zhang2025memgen} explores generative latent memory, where memory is woven into a continuous latent space that evolves with experience. Related systems such as ReMe~\cite{cao2025remember} and Generative Agents~\cite{park2023generative} further highlight the role of structured memory in supporting long-term reasoning and consistency. These works view memory evolution as an organizational process.

\paragraph{Memory self-distillation}
Those works aim to transform episodic experience into reusable knowledge, skills, or workflows. ExpeL~\cite{zhao2024expel} extracts abstract insights and rules from interaction trajectories to guide future behavior. AWM~\cite{wang2024awm} stores reusable task workflows, enabling agents to transfer procedural knowledge across tasks. MUSE~\cite{yang2025learning} further organizes memory hierarchically to accumulate strategic and procedural competence over long horizons. Collectively, these works treat memory evolution as a distillation process that converts episodic experiences into a persistent capability.

\subsection{3.2 Task Self-switching}
\label{sec:task_self_switching}
Existing works on task self-switching can be broadly categorized into two classes: task self-selection and task self-generation. These two lines differ in whether tasks are adaptively selected from an evolving candidate set or generated online conditioned on the agent’s self-state and environment dynamics.

\paragraph{Task self-selection}
Those works formulate task switching as a selection or scheduling problem, where an agent dynamically chooses which task to pursue next based on learning progress or interaction feedback. Representative works include Self-Evolving Curriculum (SEC)~\cite{chen2025self}, which learns adaptive task scheduling policies; WebRL~\cite{qi2024webrl}, which employs a self-evolving online curriculum for web interaction; Agent0~\cite{xia2025agent0}, which co-evolves a curriculum agent with an executor; as well as hierarchical embodied systems such as Mobile-Agent-E~\cite{wang2025mobile}, where high-level managers perform goal or subgoal selection during execution.

\paragraph{Task self-generation}
Those works focus on creating new tasks online rather than selecting from a predefined pool, enabling task switching driven by the agent's self-state and exploration demands. ZeroGUI~\cite{yang2025zerogui} automatically generates GUI tasks from interface states to support online learning; AgentEvolver~\cite{zhai2025agentevolver} generates new tasks via self-questioning to guide exploration in novel environments; and WebEvolver~\cite{fang2025webevolver} leverages a coevolving world model to generate self-instructed interaction tasks.

\subsection{3.3 Environment Self-prediction}
\label{sec:env_self_prediction}
Existing works on environment self-prediction can be broadly categorized into three classes: understanding World Models (WMs), generative world models, and collaborative world models. Understanding world models learn predictive latent representations that support predicting and planning without necessarily decoding raw observations. Generative world models explicitly model the observation-generation process (e.g., next-frame or rollouts) with probabilistic generative models, while collaborative world models integrate multiple predictive modules to support action-conditioned, uncertainty-aware, and long-horizon environment prediction.

\paragraph{Understanding world models}
Those works focus on learning predictive latent representations of environment dynamics, typically instantiated via RSSM-style latent dynamics models~\cite{hafner2025mastering,hafner2025training} or JEPA-style predictive representation learning~\cite{assran2023self,assran2025v}. These works can achieve environment self-prediction in self-evolving embodied AI by continually evolving latent representations under distribution shifts, e.g., continual world-model updates for long-horizon tasks in EvolvingAgent~\cite{feng2025evoagent}; task-aware world modeling for navigation in ModularAgent~\cite{zhan2026modularagent}.

\paragraph{Generative world models}
Those works explicitly predict future environment observations, commonly implemented via autoregressive Transformers or diffusion models. Recent large-scale generative world models, such as Genie~\cite{bruce2024genie}, Genie-2~\cite{parker2024genie}, and Matrix-Game-2~\cite{he2025matrix}, learn general-purpose environment dynamics from large-scale video data and enable long-horizon imagination. In Minecraft and related open-world environments, interactive generative world models such as MineWorld~\cite{guo2025mineworld} and MineDreamer~\cite{zhou2024minedreamer} provide action-conditioned video generation, and OA~\cite{liu2025continual} plans with online world models in continual RL, forming a practical foundation for environment self-prediction when combined with continual data collection and online refinement.

\paragraph{Collaborative world models}
Those works emphasize how multiple world models are composed, retrieved, aligned, or co-evolved to support environment self-prediction. Recent works have explored several forms of world-model collaboration. WorMI~\cite{yoo2025world} incorporates independently learned domain-specific world models into an embodied agent through test-time composition, enabling cross-domain adaptation by retrieving and aligning relevant world-model knowledge. COMBO~\cite{zhang2025combo} learns compositional world models for embodied multi-agent cooperation, where partial egocentric observations and multiple agents' actions are integrated to simulate joint future dynamics for online planning. WebEvolver~\cite{fang2025webevolver} introduces a co-evolving World Model that collaborates with the agent policy by generating self-instructed training data and providing look-ahead imagination during inference. WKM~\cite{qiao2024agent} further shows that multiple world models can provide both global prior and local dynamic knowledge to facilitate agent planning.

\subsection{3.4 Embodiment Self-adaptation}
\label{sec:emb_self_adaptation}
Existing works on embodiment self-adaptation can be broadly categorized into three classes: embodiment self-transfer, embodiment self-calibration, and embodiment self-recovery. This software-level component adapts internal representations to changes in the physical embodiment without modifying the hardware itself.

\paragraph{Embodiment self-transfer}
Those works transfer software policies or representations across different given physical embodiments. Representative works include GET-Zero~\cite{patel2025get}, which models embodiment topology for zero-shot generalization; Body Transformer (BoT)~\cite{sferrazza2024body}, which incorporates embodiment structure into policy learning; and PEAC~\cite{ying2024peac}, which pretrains transferable representations for cross-embodiment control.

\paragraph{Embodiment self-calibration}
Those works target continuous changes in actuation limits and physical constraints, such as mass variation, friction, contact dynamics, and sensing noise. This line typically relies on online system identification or latent parameter estimation to recalibrate internal embodiment models during execution. A classical foundation is the universal policy with online system identification (UP-OSI)~\cite{yu2017preparing}, with recent extensions enabling active or continual calibration in contact-rich scenarios, such as SPI-Active~\cite{sobanbabu2025sampling} and Online Friction Coefficient Identification (OFCI)~\cite{kim2025online} for legged robots.

\paragraph{Embodiment self-recovery}
Those works address sudden degradation or failure of sensing, actuation, or computational resources, requiring rapid restoration of executable behavior. This line emphasizes fault-aware adaptation and recovery rather than parameter refinement. Representative works include quality-diversity based damage recovery~\cite{allard2023online}, which maintains diverse behavioral repertoires for fast adaptation under damage, and learning-based fall or failure recovery policies for legged robots~\cite{yang2023learning}.

\subsection{3.5 Model Self-evolution}
\label{sec:model_self_evolution}
Existing works on model self-evolution can be broadly categorized into three classes: model self-restructuring, model self-optimization and model self-evaluation.

\paragraph{Model self-restructuring}
Those works evolve the internal architecture of models through modular growth, expert composition, or adaptive routing to meet changing requirements across embodiments, environments, and tasks. Representative works include self-supervised architecture search via masked distillation~\cite{yan2024masked}, adaptive Mixture of Expert (MoE) models for continual learning~\cite{jung2024pmoe, yu2024boosting}, dynamic adapter or LoRA expert composition~\cite{ge2025dynamic,araujo2024learning}, and modular growth with expert pruning~\cite{liu2024efficient}.

\paragraph{Model self-optimization}
Those works focus on autonomously improving training and parameter update strategies through continual tuning, self-generated feedback, or test-time learning. Typical works include continual instruction tuning without human labels~\cite{kang2025self}, learning from language feedback in SRT~\cite{hu2024teaching}, adversarial self-improvement in SEAS~\cite{diao2025seas}, test-time reinforcement learning~\cite{zuo2025ttrl}, and iterative self-refinement and reflection-based optimization~\cite{madaan2023self,shinn2023reflexion}. Within test-time learning, EATA-C~\cite{tan2025uncertainty} shows that reliable test-time self-adaptation depends on selecting trustworthy unlabeled samples, pioneering entropy-based criteria to filter ambiguous or harmful inputs for self-optimization.

\paragraph{Model self-evaluation}
Those works evolve evaluation criteria and reward signals to guide learning in the absence of reliable ground-truth supervision. This line is characterized by LLM-as-a-judge frameworks~\cite{zheng2023judging}, self-evolved reward learning~\cite{huang2024self}, reinforcement learning from meta-evaluation~\cite{rentschler2026rlme}, learning without external rewards~\cite{zhao2025learning}, and multi-judge co-evolution~\cite{chen2025multi}.

\subsection{3.6 Physical Embodiment Evolution}
\label{sec:Physical_Embodiment_Reconfiguration}
Existing works on physical embodiment evolution mainly focus on morphological reconfiguration and functional reconfiguration. The former changes the agent’s physical form, while the latter changes its physical sensing or actuation capabilities.

\paragraph{Morphological reconfiguration}
Those works change the embodiment shape or mechanical organization of an agent. EmbeddedSM~\cite{sun2023embedded} changes embodiment and limb geometry through integrated actuation and shape locking; physical polygon meshing (PhysicalPM)~\cite{belke2023morphological} changes module geometry and connectivity; shape-changing tensegrity blocks (SCTB)~\cite{zhao2025modular} combine deformation and modular assembly to construct different physical structures; and HEXEL \cite{yoder2024hexagonal} rearranges electrohydraulic modules through reversible magnetic connections to construct robotic systems with different physical morphologies and locomotion configurations.

\paragraph{Functional reconfiguration}
Those works change physical functions by replacing, rearranging, or repurposing sensing and actuation components. M4~\cite{sihite2023multi} repurposes appendages as wheels, legs, hands, or thrusters; UniCore~\cite{li2025reciprocal} combines a shared actuation core with interchangeable limbs; and FlexiFly~\cite{zhao2025flexifly} attaches task-relevant sensing and actuation modules under software guidance.

\section{4. Applications}
\label{sec:applications}
This section discusses three practical applications of self-evolving embodied AI: autonomous service robotics, autonomous driving, and autonomous Unmanned Aerial Vehicles (UAVs). A case study further illustrates the co-evolution between software and hardware.

\subsection{4.1 Autonomous Service Robotics}
\label{sec:app_service}
Autonomous service robots operate in open-ended human-centered environments where object layouts, user preferences, and task specifications change over time~\cite{zhang2025effective}. Recent advances in open-vocabulary mobile manipulation and generalist robot policies demonstrate strong generalization across tasks and scenes~\cite{yenamandra2023homerobot, huang2025enerverse, driess2023palm}, but comprehensive deployment still requires self-evolution beyond static task execution.

In this scenario, the priority software-level components are memory self-updating, task self-switching, and embodiment self-adaptation. Memory self-updating retains user preferences and failure cases, task self-switching adjusts goals under changing instructions, and embodiment self-adaptation adjusts software policies across heterogeneous platforms. Physical embodiment evolution may involve reconfiguring manipulators, sensors, or mobility modules, with the resulting hardware changes providing feedback for subsequent software evolution. Existing studies still rely heavily on fixed tasks, short-term memory, and platform-specific policies, leaving long-horizon adaptation and software-hardware co-evolution insufficiently explored.

\subsection{4.2 Autonomous Driving}
\label{sec:app_driving}
Autonomous driving represents a safety-critical instantiation of embodied AI~\cite{yurtsever2020survey} in dynamic environments. Real-world traffic exhibits non-stationary dynamics, rare corner cases, and complex interactions with humans~\cite {schwarting2018planning}, which cannot be exhaustively captured during training. EPMF~\cite{tan2024epmf} attributes perception failures to modality discrepancies and proposes discrepancy-aware modality selection and fusion, improving the reliability of multi-sensor 3D semantic segmentation. Recent work increasingly relies on predictive world models~\cite{peng2025navigscene} and high-level reasoning~\cite{sun2025sparsedrive} to support planning and simulation. 

In this scenario, the priority software-level components are environment self-prediction, model self-evolution, and embodiment self-adaptation. Environment self-prediction anticipates future trajectories and risks, model self-evolution maintains robustness under distribution shifts, and embodiment self-adaptation handles sensor calibration, vehicle dynamics, and hardware degradation. In safety-critical settings, physical embodiment evolution should be restricted to validated changes in sensing or actuation configurations, with corresponding software recalibration. Existing research is still largely evaluated on static datasets or closed benchmarks, often emphasizing short-horizon prediction rather than long-horizon evolution.

\subsection{4.3 Autonomous UAVs}
\label{sec:app_uav}
Autonomous UAVs \cite{chang2023review, feng2024u2udata} operate under fast dynamics, partial observability, and strict physical constraints, making them a natural testbed for self-evolving embodied AI. UAV missions such as inspection \cite{li2023design}, planning \cite{deng2025autonomous}, and search-and-rescue \cite{feng2025u2udata} require continual adaptation to dynamic environments \cite{wang2019autonomous}, weather conditions \cite{feng2024multi}, and task objectives~\cite{du2025survey}.

In this scenario, the priority software-level components are embodiment self-adaptation, environment self-prediction, and memory self-updating. For modular UAVs~\cite{zhao2025flexifly}, physical embodiment evolution may involve reconfiguring payload, sensing, or actuation modules, while the software adapts its perception and control policies to the resulting hardware configuration. Environment self-prediction supports wind-field estimation and obstacle-motion prediction, while memory self-updating retains rare but safety-critical experiences. These processes are constrained by onboard computation, battery capacity, communication bandwidth, and real-time control requirements. Existing studies often assume fixed physical embodiments or simulation-only evaluation, limiting the verification of long-horizon software-hardware co-evolution.

\subsection{4.4 Case Study}
Consider a home service robot asked to ``prepare a cup of tea for the user'' in a changing household environment. The agent first performs memory self-updating by retrieving and updating user preferences, object locations, and past failure cases~\cite{park2023generative,xu2025amem}. Then, task self-switching decomposes the goal into feasible sub-tasks, such as finding the kettle, checking water, grasping the cup, and delivering it to the user, based on language-grounded affordances and open-ended task exploration~\cite{chen2025self, xia2025agent0, zhai2025agentevolver}. Meanwhile, environment self-prediction estimates future states and action consequences, such as whether opening a cabinet may reveal the target cup or whether a path will be blocked~\cite{hafner2025mastering,fang2025webevolver}. During execution, embodiment self-adaptation adjusts software policies according to the robot’s physical constraints~\cite{patel2025get, sferrazza2024body, ying2024peac}. If software adaptation is insufficient, physical embodiment evolution can replace an end effector or sensing module. Finally, model self-evolution refines policies, reasoning procedures, or action representations from feedback and failures~\cite{jung2024pmoe, yu2024boosting, liu2024efficient}. The five software-level components co-evolve with each other, while hardware reconfiguration and subsequent software adaptation form a software-hardware co-evolution loop.

Current systems mainly evaluate predefined task completion rather than long-horizon co-evolution among software-level components and reconfigurable hardware~\cite{yenamandra2023homerobot}. Most methods still depend on fixed physical embodiments and improve isolated capabilities, making controllable software-level and software-hardware co-evolution difficult to verify.

\section{5. Future Directions}
\label{sec:future}
Self-evolving embodied AI presents a new paradigm for moving embodied AI beyond human-crafted setting to in-the-wild setting. Future research should focus on the controllability, trustworthiness, and swarm-level systems.

\subsection{5.1 Controllable Self-evolution}
A core challenge for self-evolving embodied AI is ensuring that the self-evolution process remains controllable under real-world constraints. Uncontrolled software evolution or physical embodiment evolution may cause unstable updates, unsafe actions, or unintended changes in the agent’s objectives or physical configuration.

At the action level, evolved policies should satisfy collision, actuation, energy, and task-specific safety constraints. At the component level, software evolution should be regulated by trigger thresholds and adaptation budgets, while physical embodiment evolution should require feasibility and safety checks. At the system level, fast perception-action control should be separated from slower software evolution and physical embodiment evolution. Quantifiable criteria such as stability margins and risk bounds will be essential for predictable operation.

\subsection{5.2 Trustworthy Self-evolution}
The applications of self-evolving embodied AI to real-world systems require self-evolution to be safe, transparent, accountable, and aligned with human values. Continuous monitoring and verification should ensure that evolved software and evolved physical embodiments do not produce unsafe or misaligned behavior.

Software updates and physical embodiment evolution should be evaluated through simulation, hardware-in-the-loop testing, uncertainty estimation, and safety monitoring. Evolution logs should record both software and hardware changes, while rollback mechanisms should restore safe software and hardware configurations when failures occur.

Human-AI alignment should be embedded into self-evolution rather than treated as an external constraint. Future work should incorporate human oversight, preference constraints, and approval mechanisms for high-risk changes. In safety-critical applications, changes to task objectives, model structures, or physical embodiments should require stronger verification and human authorization.

\subsection{5.3 Swarm Self-evolution}
While existing studies primarily focus on individual agents, many applications involve multiple agents operating collectively. Swarm self-evolution requires agents to share experiences, coordinate software evolution, and allocate or reconfigure hardware functions under resource constraints. Future research should explore distributed memory, policy sharing, role differentiation, and hardware-resource allocation. Understanding emergent behaviors and aligning individual evolution with swarm-level objectives are also crucial. Collective verification is necessary to prevent local software or hardware changes from causing globally unsafe behavior.

\section{6. Conclusion}
This paper presents self-evolving embodied AI as a new paradigm for moving embodied AI from human-crafted setting toward in-the-wild setting. It comprises five software-level components and a hardware-level component for physical embodiment evolution. Co-evolution among software-level components and software-hardware co-evolution jointly support continually adaptive intelligence with autonomous evolution. This framework incorporates hardware into self-evolving embodied AI while distinguishing physical embodiment evolution from software-level embodiment self-adaptation. We believe that self-evolving embodied AI can provide a new perspective toward general artificial intelligence.

\section{FUNDING}
This work is supported by the  Beijing National Research Center for Information Science and Technology under Grant No.BNR2026TD03005.

\section{AUTHOR CONTRIBUTIONS}
T.F., X.W., and W.Z. conceived the work. T.F., X.W., and W.Z. contributed to the definition, framework, components, mechanisms, and roadmap of self-evolving embodied AI. T.F. and X.W. contributed to the review of state-of-the-art works for realized components. T.F., X.W., and W.Z. contributed to the writing of this paper. W.Z. supervised the whole project.

\paragraph{\textbf{Conflict of interest statement}} None declared.

\bibliographystyle{nsr}
\bibliography{SEAI}

\end{document}